\documentclass
[preprintnumbers,reprint,prd,amsmath,amssymb,nofootinbib]
{revtex4-1}

\usepackage{multirow}
\usepackage{bbm}
\usepackage{graphicx}
\usepackage[tight]{subfigure}

\usepackage{color}

\newcommand{\ptitle}[1]{\emph{#1}}

\newcommand{\diag}{\mathrm{diag}}
\newcommand{\meg}{\mu \rightarrow e \gamma}
\newcommand{\order}{\mathcal{O}}
\newcommand{\eV}{\ \mathrm{eV}}
\newcommand{\GeV}{\ \mathrm{GeV}}
\newcommand{\TeV}{\ \mathrm{TeV}}
\newcommand{\TUD}{Institut f\"{u}r Kern- und Teilchenphysik, TU Dresden,
  01069 Dresden, Germany}

\begin{document}

\title{Lepton flavour violation from right-handed neutrino thresholds}
\author{Jae-hyeon~Park}
\altaffiliation{Present address:
Departament de F\'{i}sica Te\`{o}rica and IFIC,
Universitat de Val\`{e}ncia-CSIC,
46100, Burjassot, Spain}
\affiliation{\TUD}

\begin{abstract}
Charged lepton flavour violation is reappraised in the context of
supersymmetric see-saw mechanism.
It is pointed out that
a non-trivial flavour structure of right-handed neutrinos,
whose effect has been thus far less studied,
can give rise to significant
slepton flavour transitions.
Under the premise that 
the neutrino Yukawa couplings are of $\order(1)$,
the right-handed neutrino mixing contribution
could form a basis of the
$\meg$ amplitude
which by itself might lead to an experimentally accessible rate,
given a typical low-energy sparticle spectrum.
Emphasis is placed on
the crucial role of the recently measured
lepton mixing angle $\theta_{13}$ as well as
the leptonic $CP$-violating phases.
\end{abstract}
\maketitle



The see-saw mechanism is
one of the most compelling explanations for the tininess of neutrino masses
\cite{see-saw}.
The supersymmetric
type-I see-saw model (see e.g.\ \cite{Strumia:2006db}),
arguably the prototypical incarnation of this mechanism,
includes the following
terms in the superpotential,
\begin{equation}
  \label{eq:W}
  \Delta W =
    H_d L\, Y_e e^c
  - H_u L\, Y_\nu \nu^c
  - \frac{1}{2} \nu^c M_{\nu^c} \nu^c ,
\end{equation}
where the superfields
$H_d$ and $H_u$ are the down- and the up-type Higgs doublets,
$L$ is the SU(2) doublet lepton,
and $e^c$ and $\nu^c$ are the charged and the neutral
SU(2) singlet leptons, respectively.
In this article, it shall be assumed that
the leptons (including $\nu^c$) come in three flavours and
that $L$ and $e^c$ are in the basis where $Y_e$ is diagonal and positive.
The basis of $\nu^c$ shall be switched as is convenient.
Suppose that the eigenvalues of $M_{\nu^c}$ are much larger than
the weak scale.
At a low energy scale,
this underlying theory then leads to the effective neutrino mass matrix,
\begin{equation}
  \label{eq:see-saw}
  m_\nu = v_u^2\, Y_\nu M_{\nu^c}^{-1} Y_\nu^T = U^* \widehat{m}_\nu U^\dagger ,
\end{equation}
where $v_u$ is the vacuum expectation value of $H_u$ and
$U$ is the Pontecorvo-Maki-Nakagawa-Sakata lepton mixing matrix \cite{PMNS}.
A hatted matrix is diagonal containing
the non-negative eigenvalues.
The essence of the see-saw mechanism is that
one can have $\widehat{m}_\nu \lesssim \mathrm{eV}$ without
introducing a Yukawa coupling unnaturally smaller than $\order(1)$,
thanks to the suppression by $v_u M_{\nu^c}^{-1}$.

\begin{figure}[b]
  \centering
  \includegraphics{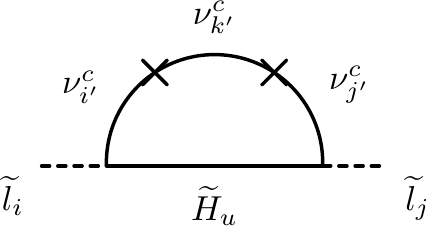}
  \caption{Schematic Feynman graph depicting the right-handed neutrino
    loop correction to the slepton mass matrix.
    The primed and the unprimed indices are in the bases
    where $Y_\nu$ and $Y_e$ are diagonal, respectively.
    The corresponding scalar loop is understood to be taken into account.}
  \label{fig:neutrino loop}
\end{figure}

This model is known to enhance lepton flavour violation
possibly to a measurable extent \cite{Borzumati:1986qx}.
Even if one imposes a flavour-blind boundary condition on the slepton
mass matrices at some high scale, $M_X$,
the renormalization group running down to a lower scale
causes flavour-violating slepton mass terms.
A graphical illustration of this effect is
Fig.~\ref{fig:neutrino loop}.
This diagram changes a slepton of flavour $i$ to flavour $j$.
A closer look at the loop reveals that
there are two different sources of flavour conversion:
the vertices and the crosses designating
right-handed neutrino mass insertions.
The former has been the focus of attention
in the majority of the literature to date.
For a review of such studies,  
see e.g.\ \cite{Masiero:2004js}.
By contrast, the latter has been less considered. 
To put it in a nutshell,
the main aim of this article is to prove the capacity of
the latter source that might 
lead to rich flavour physics in the lepton sector.

This is not the first analysis of lepton flavour violation
with right-handed neutrino mixing taken into account
(see e.g.\ \cite{Blazek:2001zm,Ellis:2001yza,Ibarra:2009bg,Alonso:2011jd}).
To the author's knowledge,
Ref.~\cite{Blazek:2001zm} is the first study in which
$\nu^c$ are sequentially decoupled.
However,
they do not state any implication of this treatment
for lepton flavour violation.
In Ref.~\cite{Ellis:2001yza},
they mention enhancement of $\meg$ due to non-degeneracy of $\nu^c$.
The decay rate does grow as
the non-degeneracy parameter $\varepsilon_N$ in their article decreases
from unity towards zero.
This change, however, is the effect from
both the left- and the right-handed mixings,
implied by $Y_\nu$ assumed to be symmetric.
This makes it difficult to disentangle the net non-degeneracy effect.
Indeed, the above variation decreases
the overall $\nu^c$ mass scale at the same time.
This increases the logarithm multiplying the degenerate contribution as well.
Furthermore, one can check numerically that
$\meg$ is enhanced by a similar factor for small $\varepsilon_N$
even if the right-handed mixing is switched off.
Most notably,
the authors of Ref.~\cite{Ibarra:2009bg}
estimate rates of lepton flavour violating decays
arising purely from right-handed mixing.
The present work differs from theirs at least in the following two points.
First, they assume a hierarchical structure of $Y_\nu$ whilst
its eigenvalues are assumed to be of similar magnitudes here.
This makes the two scenarios complementary to each other.
Second, possibly due to the Yukawa structure,
they do not relate $\theta_{13}$
to lepton flavour violation.
In a (semi) effective theory approach,
lepton flavour violation with universal $\widehat{Y}_\nu$
is analysed in Ref.~\cite{Alonso:2011jd}.
Focusing on this case,
they do not mention the relevance of the Majorana phases
which play an interesting role when $\widehat{Y}_\nu$
departs from the universal form
as shall be presented in this work.

It should be appropriate to begin by reviewing
the renormalization of slepton mass matrices.
Flavour-changing correction is
most significant in the left-handed slepton
mass matrix, which reads in the leading log approximation
\cite{Borzumati:1986qx,Masiero:2004js},
\begin{equation}
  \label{eq:k-dependent leading log}
  \Delta (m^2_{\widetilde{L}})_{ij} = 
  \frac{-1}{8\pi^2} (3 m^2_0 + A_0^2)
  H_{ij}
,
\end{equation}
where $m_0$ and $A_0$ are the universal soft mass
and trilinear coupling of sleptons at $M_X$, respectively.
The flavour structure of this correction is dictated by the hermitian matrix
\cite{Ellis:2002fe},
\begin{equation}
\label{eq:H}
  H_{ij} \equiv (Y_\nu)_{ik}^* (Y_\nu)_{jk} \ln \frac{M_X}{M_k} .
\end{equation}
Repeated indices are summed over.
The three eigenvalues of $M_{\nu^c}$ are denoted by $M_k$
which demands that $Y_\nu$ be in the basis where
$M_{\nu^c}$ is diagonal.

In this basis,
one can decompose $Y_\nu$ in the form,
\begin{equation}
  \label{eq:svd Yn}
  Y_\nu = (U^l)^*\, \widehat{Y}_\nu V ,
\end{equation}
where each of the two unitary matrices
describes an eigenstate mismatch between a pair of matrices.
Analogous to the Cabibbo-Kobayashi-Maskawa matrix
originating from the misalignment between $Y_d$ and $Y_u$ in the quark sector,
$U^l$ is the twist between the left-handed rotations of $Y_e$ and $Y_\nu$.
On the other hand,
$V$ describes the mismatch between
$Y_\nu$ and $M_{\nu^c}$.
In the basis where $\nu^c$ instead diagonalizes $Y_\nu$,
one would thus find $M_{\nu^c} = V^* \widehat{M}\, V^\dagger$.
Obviously, there is no quark-sector analogue of $V$
in the Standard Model.

For the following discussions, it is convenient to split~(\ref{eq:H})
into the two parts,
\begin{equation}
  \label{eq:H split}
  H = H_\mathrm{D} + H_\mathrm{ND} ,
\end{equation}
of which
\begin{equation}
  \label{eq:HD}
  H_\mathrm{D} = Y_\nu^* Y_\nu^T \ln \frac{M_X}{M} =
  U^l \widehat{Y}_\nu^2 (U^l)^\dagger \ln \frac{M_X}{M} ,
\end{equation}
is indifferent to the individual masses of
the right-handed neutrinos whilst
\begin{equation}
  \label{eq:HND}
  (H_\mathrm{ND})_{ij} = (Y_\nu)_{ik}^* (Y_\nu)_{jk} \ln \frac{M}{M_k} ,
\end{equation}
arises from their non-degeneracy.

The ``degenerate'' contribution (\ref{eq:HD}), is due to the running
from $M_X$ down to $M$,
the overall scale of the right-handed neutrino masses.
This is the result that one would obtain by
decoupling the three right-handed neutrinos at $M$ at the same time.
The only source of flavour violation therein is $U^l$,
since it discards
the $k$-dependence of the logarithm in~(\ref{eq:H}),
thereby letting $V$ drop out.
This endows this contribution with a notable property:
no left-handed mixing implies
no lepton flavour violation.
This is often summarized in the formulation:
$l_i\rightarrow l_j\gamma$ is governed by $(Y_\nu Y_\nu^\dagger)_{ij}$
(see e.g.\ \cite{Casas:2001sr}).
There are two cases where it suffices to consider
$H_\mathrm{D}$ only.
The first is the trivial case
where the right-handed neutrinos are really degenerate.
By setting $M$ to the common $M_k$,
one can trivially eliminate the contribution
from $H_\mathrm{ND}$ in~(\ref{eq:HND}).
The second is when $\widehat{Y}_\nu$ is so hierarchical that
the largest entry dominates the loop correction
(see e.g.\ \cite{Moroi:2013vya}).
Note from~(\ref{eq:H}) and~(\ref{eq:svd Yn}) that
$(Y_\nu)_{ik}^* (Y_\nu)_{jk} \ln M_k \approx
(U^l)_{i3} (U^l)_{j3}^*
\widehat{Y}_{\nu,3}^2
\,|V_{3k}|^2 \ln M_k$,
where $\widehat{Y}_{\nu,3}$ denotes the largest element in
$\widehat{Y}_\nu$.
With $M$ chosen so that $\ln M = |V_{3k}|^2 \ln M_k$,
one can easily verify that $H_\mathrm{ND}$ vanishes.
One can interpret this derivation
within Fig.~\ref{fig:neutrino loop}.
The strength of the $\widetilde{H}_u$--$\nu^c_{j'}$--$\widetilde{l}_j$
coupling
is given by $[U^l\widehat{Y}_\nu]_{jj'}$.
A mass insertion converting $\nu^c_{i'}$ to $\nu^c_{k'}$
stems from $(M_{\nu^c})_{k'i'}$.
If $\widehat{Y}_{\nu,3}$ is dominant,
one can fix the internal indices
$i'$ and $j'$ to 3, which isolates the effect of mass insertions
from slepton flavour changes.
This means that the loop will conserve slepton flavours
to a good approximation
in the limit where $U^l = \mathbbm{1}$,
which coincides with how (\ref{eq:HD}) acts.

This observation also provides a hint on
under which circumstances $H_\mathrm{ND}$ becomes relevant, i.e.\
when the mass insertions on the $\nu^c$ line
(named the second source of flavour violation
in the introductory text)
cause significant slepton flavour changes.
To see this, it is useful to imagine an extreme case
where flavour violation from the first source has been eliminated,
i.e.\ $U^l = \mathbbm{1}$.
In this limit, $i = i'$ and $j = j'$ on the two vertices
in Fig.~\ref{fig:neutrino loop}, whose strengths are then given by
$\widehat{Y}_{\nu,i}$ and $\widehat{Y}_{\nu,j}$, respectively.
In order to find an interesting effect in the
$\widetilde{\mu}$--$\widetilde{e}$ transition for instance,
one needs to assume that neither $\widehat{Y}_{\nu,1}$ nor
$\widehat{Y}_{\nu,2}$ is highly suppressed.
In view of the rationale behind the see-saw mechanism,
one might advocate this all-$\order(1)$ structure of
$\widehat{Y}_{\nu}$,
arguing that they are the most natural orders of magnitude they could have.
Leaving aside contemplation of Yukawa textures,
the above pattern of $\widehat{Y}_{\nu}$ shall be employed
throughout what follows
as a logical possibility that best manifests the effect
of a non-trivial flavour structure of $M_{\nu^c}$.

Next, 
$H_\mathrm{ND}$ in~(\ref{eq:HND}) is to be scrutinized.
For this, one can split each Yukawa eigenvalue in the form,
$\widehat{Y}_{\nu,k} = y_\nu + \Delta_k$,
where the $k$-independent piece, $y_\nu$, is supposed to
have a value characteristic of the three components,
and $\Delta_k$ is a ``perturbation'' around this,
whose magnitude may or may not be much smaller than $y_\nu$.
With the aid of the see-saw formula (\ref{eq:see-saw}), and
the decomposition~(\ref{eq:svd Yn}) with the eigenvalues
split as prescribed above,
one can extract the following contribution from $H_\mathrm{ND}$:
\begin{equation}
  \label{eq:HNDzero}
  (H_\mathrm{ND}^0)_{ij} =
  \frac{1}{2} \frac{M^2}{y^2_\nu\, v_u^4}
  \,{{m}_{k}^2}
  U_{ik} U^*_{jk}
,
\end{equation}
making use of the approximation,
$\ln (M/M_k) \approx (M^2 / M_k^2 - 1)/2$,
where $m_k$ denotes the $k$-th diagonal entry of $\widehat{m}_\nu$.
This expression captures the qualities of the contribution
from the right-handed mixing,
provided that all the entries of $\widehat{Y}_\nu$ are of the same order.
The error due to the above approximation of log as well as
terms depending on $\Delta_k$, which may or may not be negligible,
comprise the rest of $H_\mathrm{ND}$,
which is formally at higher orders
than $H_\mathrm{ND}^0$, in powers of $(M^2/M_k^2-1)$ or $\Delta_k$.

There is a nice interpretation of (\ref{eq:HNDzero})
within a formulation using spurions (see e.g.\ \cite{Alonso:2011jd}).
Since $H_\mathrm{ND}^0$ is the contribution from the universal part of
$\widehat{Y}_\nu$, one can consider a subgroup of flavour symmetry
which leaves (\ref{eq:W}) invariant, under the condition that
${Y}_\nu = y_\nu \mathbbm{1}$.
The small charged lepton Yukawas shall be ignored.
One such group is $\mathrm{SU}(3)_{L-\nu^c}$ under which
$L$ and $\nu^c$ are $\mathbf{3}$ and $\overline{\mathbf{3}}$, respectively.
For this, one needs to promote $M_{\nu^c}$ to a spurion transforming as
a sextet under $\mathrm{SU}(3)_{L-\nu^c}$.
The flavour-violating part of $H$ in the left-handed slepton mass matrix
(\ref{eq:k-dependent leading log}) then belongs to an octet.
Now, the task is to construct octets out of $M_{\nu^c}$.
The simplest tensor product is $M_{\nu^c} M_{\nu^c}^\dagger$.
One may as well take its inverse matrix
which also transforms as an octet and multiply it by
appropriate factors for it to be part of $H$.
 From Fig.~\ref{fig:neutrino loop}, it is clear that $H$ should
pick up $y_\nu^2$.
One further needs $M^2$ in order to match the dimensions.
Combined with~(\ref{eq:see-saw}), the result then reads
\begin{equation}
 y_\nu^2 M^2 M_{\nu^c}^{-1\dagger} M_{\nu^c}^{-1} =
 \frac{M^2}{y_\nu^2 v_u^4} m_\nu^\dagger m_\nu ,
\end{equation}
which coincides with~(\ref{eq:HNDzero}) up to a factor of 2.
This is the inverse matrix of the octet contribution
$\Delta_8^{(2)}$
given in~(2.17) of Ref.~\cite{Alonso:2011jd}.
Nevertheless, this does not mean any conflict since
one could arrive at $\Delta_8^{(2)}$ by expanding
$\ln (M_k/M)$ instead of $\ln (M/M_k)$.
This would amount to taking
$M_{\nu^c} M_{\nu^c}^\dagger$,
rather than its inverse, as the starting tensor product.
Instead of being a poorer approximation of $H_\mathrm{ND}$
than $\Delta_8^{(2)}$,
$H_\mathrm{ND}^0$ has the advantage that
its flavour structure is fixed independent of the absolute neutrino masses
or their ordering, as shall be shown below.

A few comments about $H_\mathrm{ND}^0$ are in order.
First, this contribution is independent of $M_X$,
which by contrast
plays the important role of determining the overall strength
of flavour violation in $H_\mathrm{D}$.
Clearly, $H_\mathrm{D}$ goes away as one brings $M_X$ to $M$.
Second, $H_\mathrm{ND}^0$ is proportional to $M^2$.
The $M$ scale is inversely proportional to the scale of
absolute low-energy neutrino masses, as is evident from~(\ref{eq:see-saw}).
Therefore, this contribution grows as the lightest neutrino mass
decreases, for a fixed $y_\nu$.
Third, $(H_\mathrm{ND}^0)_{ij}$
would be of $\order(y_\nu^2)$ provided that 
all the components of $U$ were of $\order(1)$,
as the see-saw formula~(\ref{eq:see-saw}) suggests.
Putting this representative magnitude into~(\ref{eq:k-dependent leading log}),
one would find rough estimates of the slepton mass insertions,
$(\delta^e_{ij})_{LL} \sim 10^{-2} y_\nu^2$,
which implies that this contribution has enough potential to
reveal itself in low-energy flavour physics.
To be precise, this may be an overestimation
since there are small entries in $U$.
Fourth, the off-diagonal flavour structure of $H_\mathrm{ND}^0$
is fully controlled by those quantities that are measured
at neutrino oscillation experiments.
To see this, one can rewrite
the $k$-summation in~(\ref{eq:HNDzero}) in the form,
\begin{equation}
\label{eq:m2UU}
\begin{aligned}
  {m}_{k}^2 U_{ik} U^*_{jk} &=
  \Delta m^2_{31} U_{i3} U_{j3}^* + \Delta m^2_{21} U_{i2} U_{j2}^*
\\
 &=
  \Delta m^2_{32} U_{i3} U_{j3}^* - \Delta m^2_{21} U_{i1} U_{j1}^* ,
  \quad i \neq j,
\end{aligned}
\end{equation}
with the familiar abbreviation,
$\Delta m^2_{ij} \equiv {m}_{i}^2 - {m}_{j}^2$ \cite{pdg}.
Setting either $i$ or $j$ to 1 and
recalling that
$|\Delta m^2_{31}|$ and $|\Delta m^2_{32}|$ are about 30 times
bigger than $\Delta m^2_{21}$ \cite{pdg},
one finds that
the $\widetilde{\mu}$--$\widetilde{e}$ and
the $\widetilde{\tau}$--$\widetilde{e}$ transitions
depend strongly on $U_{13}$
whose modulus has been recently measured \cite{s13 experiment}.
On the other hand,
the $\widetilde{\tau}$--$\widetilde{\mu}$ component
is not suppressed by $U_{13}$.
It is noteworthy that (\ref{eq:HNDzero}) holds
no matter whether $U^l = U$ or not.

To appreciate the impact of right-handed neutrino mixing,
which appears only in $H_\mathrm{ND}$,
it is useful to introduce
a parameter denoted by $l$,
the ``left-handedness'' of lepton mixing.
It interpolates the two extreme cases
where the lepton mixing arises solely from
either $U^l$ or $V$ in~(\ref{eq:svd Yn}).
More concretely, it shall be chosen to be an exponent ranging between 0 and 1
that raises $U$ to the $l$-th power yielding $U^l$.
Needless to say, the above two limits amount to
$U^1 = U$ and $U^0 = \mathbbm{1}$, respectively.

The method to evaluate the low-energy slepton mass matrices
is the standard
renormalization group analysis
using one-loop beta functions (see e.g.\ \cite{Casas:2001sr}).
In this way,
one can solve the system of
three different effective field theories with
three, two, and one active $\nu^c$ flavour(s),
laid down starting from $M_X$ down to $M_1$,
below which the minimal supersymmetric standard model takes over.

First of all,
boundary conditions are imposed
on the superpotential and the soft parameters at $M_X$.
The $M_X$ scale value of $Y_e$ is obtained by integrating its
renormalization group equation from the weak scale.
The input for the neutrino-sector
consists of $\widehat{Y}_\nu$, $U^l$, and $m_k$.
The former two are chosen by hand.
The low-energy neutrino masses are specified by
the smallest of them in combination with
their ordering, for which there are two possibilities:
normal spectrum with $m_1 < m_2 < m_3$ and
inverted spectrum with $m_3 < m_1 < m_2$.
Using this input,
one can solve~(\ref{eq:see-saw}) and (\ref{eq:svd Yn})
to obtain $M_{\nu^c}$.
These $\widehat{Y}_\nu$, $U^l$, and $M_{\nu^c}$
are then used as their $M_X$ scale values.
Note that this procedure ignores the renormalization of
the dimension-5 neutrino mass operator,
which is significant at high $\tan\beta$
(see e.g.\ \cite{Antusch:2003kp,Petcov:2005yh}).
Taking it into account would amount to alteration in
the neutrino-sector boundary conditions at $M_X$.
This would qualify well as an element of a more advanced future analysis.
The soft parameters are set to the usual flavour-blind form,
i.e.\
the masses (including those of Higgses) are all $m_0$
and each trilinear coupling is
$A_0$ times the corresponding Yukawa coupling.

Then,
the renormalization group equations are solved
to result in the weak scale couplings.
For the evaluation of full $H$ in~(\ref{eq:H split}),
the three eigenstates of $M_{\nu^c}$
are integrated out
sequentially,
each at the scale equal to its own mass eigenvalue.
To isolate $H_\mathrm{D}$ for a comparison,
all the $\nu^c$ flavours are decoupled
at $M$, chosen to be the geometric mean of $M_k$.
At the weak scale,
the tree-level electroweak symmetry breaking condition
determines $\mu$ which enters into the chargino
and the neutralino mass matrices.
Its sign is chosen to be positive.

\begin{table}
  \centering
  \begin{tabular}{l|l|l}
\hline
&\multicolumn{1}{c|}{Normal ordering}&\multicolumn{1}{c}{Inverted ordering}
\\
\hline
    \multirow{2}{*}{$\nu$-sector}
& $m_1 = 0.05\eV$
& $m_3 = 0.05\eV$
    \\
& $\widehat{Y}_\nu = \diag(0.6, 0.8, 1.0)$
& $\widehat{Y}_\nu = \diag(0.8, 1.0, 0.6)$
    \\
\hline
At $M_X$
&
\multicolumn{2}{l}{
  \qquad
    $m_0 = 1\TeV$\qquad
    $A_0 = 0$
}
\\
\hline
At $M_\mathrm{weak}$
&
\multicolumn{2}{l}{
  \qquad
    $M_2 = 1\TeV$\qquad
    $\tan\beta = 20$
}
    \\
\hline
  \end{tabular}
  \caption{Default input to be used
    in the numerical analysis unless specified otherwise.
    The high scale is chosen to be
    $M_X = 2\times 10^{16} \GeV$.
    The bino mass $M_1$ is derived from $M_2$ by the unification relation.}
  \label{tab:input}
\end{table}
The numerical input for this programme is collected
in Table~\ref{tab:input}.
The scalar and the gaugino masses are chosen to be
well above the current collider bounds \cite{LHC bounds}.
To set out the lepton mixing matrix $U$,
it is convenient to follow the standard parametrization
in Ref.~\cite{pdg}.
The three mixing angles are assumed to hold their central values
from the same reference.
The remaining free parameters to be fixed are
the Dirac phase, $\delta$,
and the Majorana phases, $\alpha_{21}$ and $\alpha_{31}$ \cite{pdg}.
For illustrative purposes,
each of them is chosen to be either 0 or $\pi$,
yielding 8 combinations in total.
They are encoded as the line pattern.

\begin{figure}
  \centering
  \includegraphics{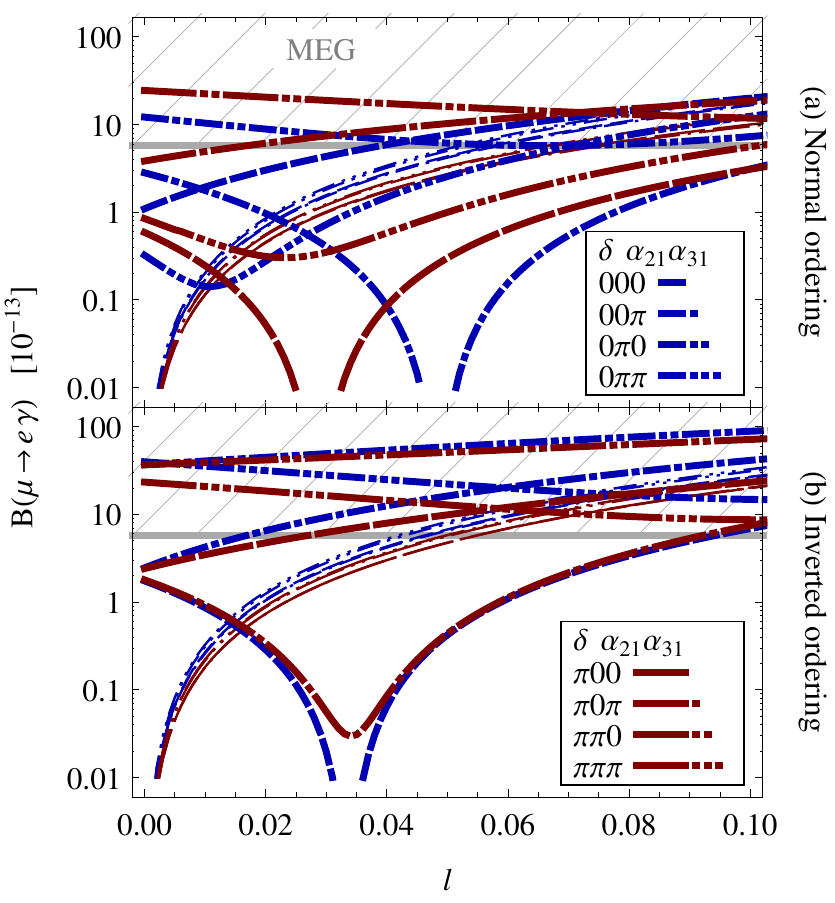}
  \caption{Branching fraction of $\meg$ as a function of
    $l$, the ``left-handedness'' of lepton mixing,
    in the case of (a) normal or (b) inverted neutrino mass ordering.
    The thin and the thick curves display
    the single and the multi-threshold results, respectively.
    Each line pattern indicates a combination of
    the Dirac and Majorana phases. 
    The upper hatched region shows the current MEG bound.}
  \label{fig:br-lhness}
\end{figure}
The above procedure leads to Fig.~\ref{fig:br-lhness},
which shows
the branching fraction of $\meg$ as a function of $l$.
Therein,
one finds an outstanding difference between
the single and
the multi-threshold results,
which are delineated by the thin and the thick curves,
respectively.
The former result corresponds to keeping only
$H_\mathrm{D}$ from~(\ref{eq:H split}).
As is clear from~(\ref{eq:HD}),
the $\meg$ rate drops to zero as
$l$ decreases, i.e.\
as the role of the left-handed mixing diminishes.
On the other hand,
the $\meg$ rate remains finite even for $l = 0$,
if one takes care of the individual $\nu^c$ thresholds,
thereby incorporating $H_\mathrm{ND}$
in~(\ref{eq:H split}).
Moreover,
those finite values might be not only of theoretical
but also of experimental interest,
since they range around the current experimental upper limit
at 90\% confidence level,
$5.7\times 10^{-13}$,
achieved by the MEG experiment \cite{Adam:2013mnn}.
It is worth highlighting that
lepton flavour violation originating purely from
the right-handed mixing
has sufficient potential to be discovered
at low-energy new physics searches.
Another point to notice is that
the branching fraction shows a drastic variation
depending on the phases,
which remains even if $l = 0$.
There are curves that intrude into the excluded region,
but they are meant to demonstrate the variety
of prediction as a function of the phases and $l$.
One could at any time resurrect a point
with a particular set of $l$ and phases,
which is ruled out by $\meg$,
by suppressing its rate.
To this end, the following modifications are helpful:
to decrease $\widehat{Y}_\nu$ or $\tan\beta$, or
to increase $m_k$, $m_0$, or the gaugino masses.

\begin{figure}
  \centering
  \includegraphics{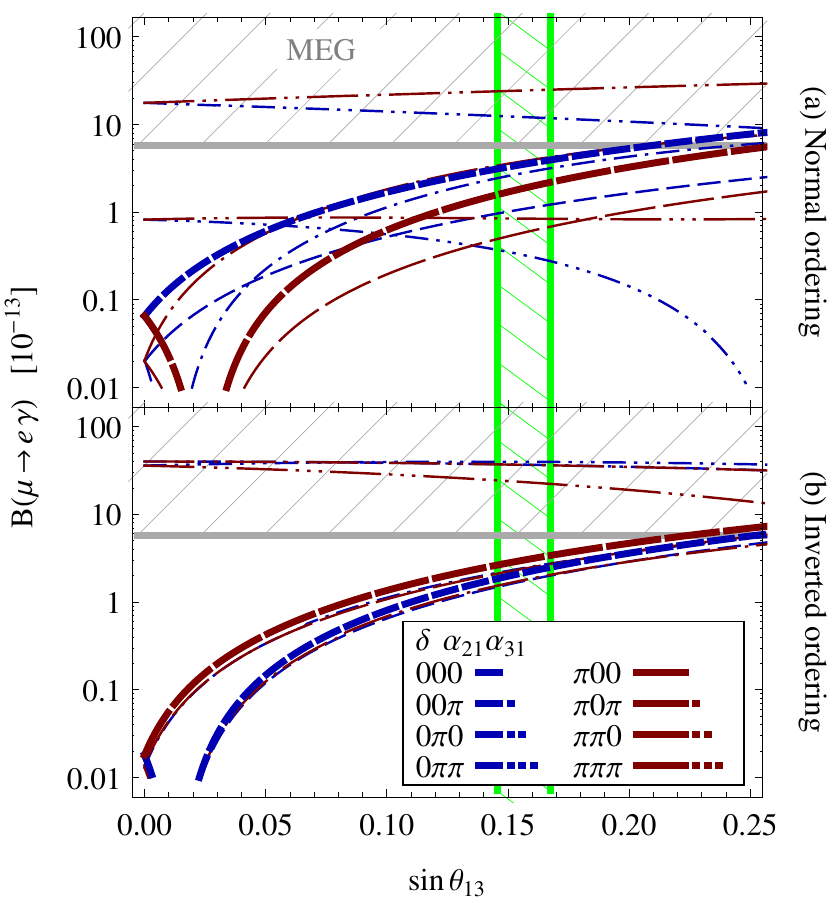}
  \caption{Branching fraction of $\meg$ for $l=0$ as a function of
    $\sin \theta_{13}$,
    in the case of (a) normal or (b) inverted neutrino mass ordering.
    The thick and the thin curves correspond to
    $\widehat{Y}_\nu = \diag(1,1,1)$ and
    the default neutrino Yukawas in Table~\ref{tab:input}, respectively.
    Each line pattern indicates a combination of
    the Dirac and Majorana phases. 
    The upper hatched region shows the current MEG bound.
    The vertical hatched strip is the measured 1 $\sigma$ range of
    $\sin \theta_{13}$.}
  \label{fig:br-s13}
\end{figure}

As explained above,
the lepton mixing angle $\theta_{13}$
plays a critical role in
lepton flavour violation from the right-handed mixing.
A graphical exposition of
this is  Fig.~\ref{fig:br-s13}.
In the plots, the decay arises solely from the right-handed mixing.
The thick curves show the properties of~(\ref{eq:HNDzero}),
which is the contribution from the
universal part of the neutrino Yukawa eigenvalues.
Indeed,
the branching fraction exhibits a strong dependence on $\sin\theta_{13}$.
Furthermore,
it is entertaining to notice that
$\meg$ is enhanced by
the departure of
$\sin\theta_{13}$ from zero.
For this uniform choice of $\widehat{Y}_\nu$,
only the Dirac phase is varied
since the Majorana phases leave no trace.
Using the standard parametrization of $U$ \cite{pdg},
one can easily verify that the Majorana phases are cancelled out
in~(\ref{eq:HNDzero}).
If one instead chooses non-universal Yukawas,
the Majorana phases turn active,
as the thin curves display.
It is then amusing to realize that
this ``Majorana effect'' belongs neither to $H_\mathrm{ND}^0$ 
nor to $H_\mathrm{D}$.
It arises from $H_\mathrm{ND} - H_\mathrm{ND}^0$.  


In summary,
it has been stressed that
the flavour structure of heavy right-handed neutrinos
could be an interesting origin of
supersymmetric lepton flavour violation.
In particular,
the mass hierarchy plus large mixing in the $\nu^c$ sector alone
might lead to sufficient $\meg$ for its experimental discovery.
When acting on
the $\widetilde{\mu}$--$\widetilde{e}$ or
the $\widetilde{\tau}$--$\widetilde{e}$ transition,
this ``right-handed'' effect is basically
controlled by $\theta_{13}$,
which renders its recent measurements far more fascinating.
Full consideration of the individual $\nu^c$ thresholds
unlocks the striking influence of the leptonic $CP$-violating phases
on flavour physics.


The author thanks Dominik St\"{o}ckinger for enlightening discussions
and a careful reading of the manuscript.
He also thanks Alejandro Ibarra for bringing Ref.~\cite{Ibarra:2009bg}
to his attention.

\end{document}